\documentclass{article}

\usepackage{arxiv}

\usepackage[utf8]{inputenc} % allow utf-8 input
\usepackage[T1]{fontenc}    % use 8-bit T1 fonts
\usepackage{url}            % simple URL typesetting
\usepackage{booktabs}       % professional-quality tables
\usepackage{amsfonts}       % blackboard math symbols
\usepackage{nicefrac}       % compact symbols for 1/2, etc.
\usepackage{float}
\usepackage{amsmath}
\usepackage{microtype}      % microtypography
\usepackage{lipsum}
\usepackage{array}% Can be removed after putting your text content
\usepackage{graphicx} 
\usepackage{subcaption}
\usepackage{multirow}
 \usepackage[
    backend=biber,
    style=nature,
    doi=true,
  ]{biblatex}

\usepackage[colorlinks=true, citecolor=blue, linkcolor=blue, urlcolor=blue]{hyperref}

\usepackage{cleveref}       % smart cross-referencing

\usepackage{doi}
\usepackage{xcolor}
\title{The overlooked need for Ethics in Complexity Science: Why it matters}

\usepackage{authblk}

\newbox{\orcid}\sbox{\orcid}{\includegraphics[scale=0.06]{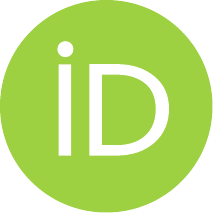}}

\author[1]{\mbox{%
	\href{https://orcid.org/0000-0001-8341-939X}{\usebox{\orcid}}\hspace{1mm}Olumide Adisa%\thanks{\texttt{your@email}}%
}}
\affil[1]{University of Suffolk- UK}

\author[2]{\mbox{%
	\href{https://orcid.org/0000-0001-7097-9400}{\usebox{\orcid}}\hspace{1mm}Enio Alterman Blay%\thanks{\texttt{your@email}}%
}}
\affil[2]{Universidade de São Paulo - Brazil}

\author[3]{\mbox{%
	\href{https://orcid.org/0000-0002-5397-0778}{\usebox{\orcid}}\hspace{1mm}Yasaman Asgari%\thanks{\texttt{your@email}}%
}}
\affil[3]{University of Zurich - Switzerland}

\author[4,5]{\mbox{%
	\href{https://orcid.org/0000-0003-2615-0712}{\usebox{\orcid}}\hspace{1mm}Gabriele Di Bona\thanks{Corresponding author: Gabriele Di Bona, email address: \texttt{gabriele.dibona.work@gmail.com}}%
}}
\affil[4]{GEMASS, Centre national de la recherche scientifique, Paris - France}
\affil[5]{Sony Computer Science Laboratories, CREF, Rome - Italy}

\author[6]{\mbox{%
	\href{https://orcid.org/0000-0003-3156-6579}{\usebox{\orcid}}\hspace{1mm}Samantha Dies%\thanks{\texttt{dies.s@northeastern.edu}}%
}}
\affil[6]{Northeastern University - USA}

\author[7]{\mbox{%
	\href{https://orcid.org/0000-0003-2409-3064}{\usebox{\orcid}}\hspace{1mm}Ana Mar\'ia Jaramillo%\thanks{\texttt{jaramillo@csh.ac.at}}%
}}
\affil[7]{Complexity Science Hub, Vienna - Austria}

\author[8]{\mbox{%
	\href{https://orcid.org/0000-0002-9238-3023}{\usebox{\orcid}}\hspace{1mm}Paulo H. Resende%\thanks{\texttt{phresende@uol.com.br}}%
}}
\affil[8]{University of Campinas - Brazil}

\author[9]{\mbox{%
	\href{https://orcid.org/0000-0002-8186-2627}{\usebox{\orcid}}\hspace{1mm}Ana Maria de Sousa Leitão%\thanks{\texttt{your@email}}%
}}
\affil[9]{INET-md, branch of FMH, University of Lisbon - Portugal}

\hypersetup{
pdftitle={\thetitle},
}

\begin{document}
\maketitle

\begin{abstract}
Complexity science, despite its broad scope and potential impact, has not kept pace with fields like artificial intelligence, biotechnology and social sciences in addressing ethical concerns.
The field lacks a comprehensive ethical framework, leaving us, as a community, vulnerable to ethical challenges and dilemmas. Other areas have gone through similar experiences and created, with discussions and working groups, their guides, policies and recommendations.
Therefore, here we highlight the critical absence of formal guidelines, dedicated ethical committees, and widespread discussions on ethics within the complexity science community. 
Drawing on insights from the disciplines mentioned earlier, we propose a roadmap to enhance ethical awareness and action. Our recommendations include (i) initiating supportive mechanisms to develop ethical guidelines specific to complex systems research, (ii) creating open-access resources, and (iii) fostering inclusive dialogues to ensure that complexity science can responsibly tackle societal challenges and achieve a more inclusive environment. 
By initiating this dialogue, we aim to encourage a necessary shift in how ethics is integrated into complexity research, positioning the field to address contemporary challenges more effectively.
\end{abstract}


\section*{Introduction}

The field of complexity science has significantly influenced a wide range of domains, from ecosystems and economies to social networks and artificial intelligence (AI). Despite its broad reach, ethical challenges in this research have received very limited attention. This oversight is particularly concerning given the ethical dilemmas faced by related fields. For instance, the rapid development and deployment of artificial intelligence have led to significant ethical issues and raised discussions about algorithmic bias, privacy violations, automated decision-making and unintended consequences~\cite{boyd2011six,zou2018design,ryan2020ai, stenseke2024computational}.

As complexity science continues to expand, the absence of a comprehensive ethical framework poses a risk of similar problems. Unlike other fields, such as AI, which have established various codes of conduct~\cite{correa2023worldwide}, recommendations~\cite{ai4good}, and laws~\cite{eu2023ai} to address ethical concerns, complexity science lacks dedicated ethical guidelines. 
As far as we know, there have been no official statements or comprehensive frameworks specifically addressing ethical questions within this field. Furthermore, the complexity science community does not have established ethical committees to oversee or address these issues, unlike its counterparts in AI and social sciences.

Thus, the need for formal guidelines and ethical discussions becomes increasingly urgent. Here, we address this gap by highlighting the necessity of integrating ethics into complexity science and suggesting a roadmap to initiate this process. Our goal is to spark discussion and encourage the development of ethical considerations within the community, drawing on insights from related fields. We hence propose a call to action for the community to prioritize these ethical concerns and work collaboratively towards a more ethically aware and responsible practice in complexity science.

\section*{Mapping ethical challenges in complexity science}

Addressing ethical concerns in complexity science is crucial due to the profound implications for both natural and human-made systems. The analysis and modeling of complex systems, i.e., systems with numerous interconnected components, nonlinear interactions and emerging behaviour, in fact pose important ethical questions in various domains. For instance, in ecosystem management~\cite{mattos2022metrics}, these models help predict the effects of environmental changes on biodiversity and habitat conservation. In genetic research, they are used to explore the potential consequences of gene editing technologies, raising ethical questions about their long-term effects. In public health, complex models track disease spread and can help to assess interventions, influencing health policies and resource allocation~\cite{dieguez2023controlling}. Social network analysis can affect our understanding of information flow and privacy among many other digital issues~\cite{BUTTS200131}. Urban planning models can shape community well-being and equity~\cite{ribeiro2021association}. Economic models used to understand financial systems can impact economic policies and risk management~\cite{mattos2022metrics}. 
This limited list of examples makes it clear how, without a structured ethical framework, there is a risk of inadvertently causing harm, exacerbating inequalities or, more in general, doing research with unethical consequences. Therefore, it is essential to evaluate the broader implications of our research and ensure that our methodologies adhere to rigorous ethical standards, safeguarding both natural environments and human societies~\cite{saltelli2020five}.

To effectively address these important ethical challenges, we must explore how different approaches to complexity can inform our ethical considerations and guide the development of a robust ethical framework.
Should the ethical principles guiding complexity studies be specifically tailored to the unique characteristics of the systems under study? Or should they follow a more generalized epistemological approach that embraces the full scope of complexity without falling into oversimplification?
This distinction is crucial. 
``Restricted complexity”---referring to the field of science focused on dynamical systems known as complex~\cite{morin2007restricted}---has led to significant advances in formalization and modeling of various processes, leading to valuable interdisciplinary insights. 
This approach, however, often remains constrained by traditional scientific paradigms that emphasize reductionism, often leading to complicated models rather than complex ones~\cite{woermann2012ethics, li2023promises}. Reductionism posits that understanding the components of a system can fully explain its overall behavior, potentially overlooking the emergent properties that arise from complex interactions among those components.
In contrast, ``generalized complexity” confronts the limitations of reductionism by focusing on the relationships between the individual parts and the broader context in which they interact~\cite{morin2007restricted}. This approach emphasizes that complex systems cannot be fully understood through reduction alone and that models must account for the interdependencies and emergent properties of systems. According to this point of view, complexity science works should think beyond the specific systems under study, and consider the ethical implications in a wider societal and environmental context, offering a broader understanding that can inform our perspectives not just as researchers, but as members of the society.

Therefore, to move forward effectively, we need to reconcile these approaches and develop a comprehensive ethical framework that both acknowledges the specificities of different systems and embraces the overarching principles of complexity science. This does not imply adopting a radically relativist stance, but rather being open to integrating diversity and engaging in a process of continuous revision~\cite{woermann2012ethics}. By doing so, we can create guidelines that are robust and applicable across various domains while remaining adaptable to the complexities inherent in each unique context.
Through this approach, we can build a framework that addresses both \textit{i)} ethical considerations regarding the nature and methods of complexity research and also \textit{ii)} broader ethical issues such as diversity, accessibility, representation, and privacy.

\section*{Limited research on ethics in complexity science}

While there has been some discussion surrounding the ethics of studying complex systems topics, such as societies~\cite{tubaro2021social,turner2019complexity}, violence~\cite{deppman2019scalingpropertiesfirearmhomicides}, financial markets~\cite{mortoza2017antecipaccao}, social inequalities~\cite{hodgson2003capitalism}, countries~\cite{hausmann2014atlas}, and agent-based systems simulating human behavior~\cite{anzola2022ethics,rand2007farol, schelling1971dynamic}, the conversation remains limited and fragmented.

\begin{figure}[t]
    \centering
    \includegraphics[width=\textwidth]{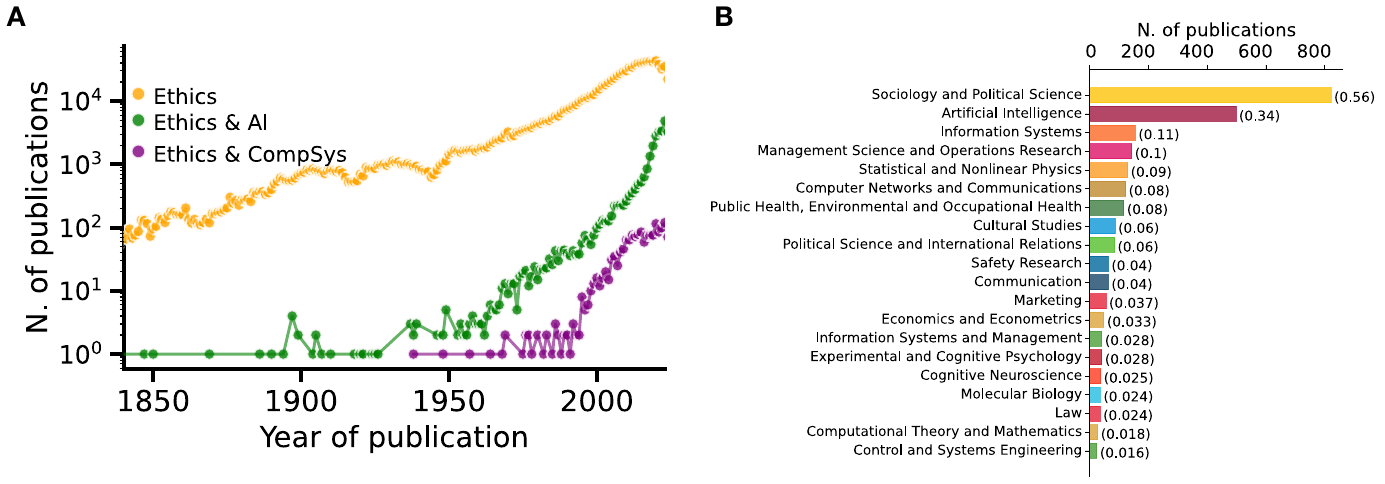}
    \caption{
        (\textbf{A}) Number of scientific papers in OpenAlex every year until 2023 with keywords referred to ``Ethics'' in yellow, to ``Artificial Intelligence'' (or AI) and ``Ethics'' in green, and to ``Complex Systems'' (or Complexity) and ``Ethics'' in purple. Complete set of keywords used for the literature search in \tableautorefname~\ref{tab:keywords_table}. 
        (\textbf{B}) Disciplinary area distribution of the papers mentioning ``Complex Systems'' and ``Ethics''. Notice how varied and interdisciplinary the selected papers are, with the biggest areas being Sociology \& Political Science and Artificial Intelligence.
    }
    \label{fig:figure1}
\end{figure}

To better quantify and understand this landscape, we conducted a bibliometric analysis using literature retrieved from the Open Alex database \cite{openalex}. 
Our preliminary findings reveal a growth in literature related to ethics and complex systems since the 1950s. However, this growth is on a smaller scale compared to the literature on ethics in artificial intelligence and general ethics (see \figureautorefname~\ref{fig:figure1}\textbf{A}). 
Moreover, the current research in this area is notably interdisciplinary, with significant contributions from social scientists. This highlights a gap in engagement from the complexity science community, which is predominantly composed of STEM researchers, and underscores the need for more dedicated resources to this topic (see \figureautorefname~\ref{fig:figure1}\textbf{B}). For further details on our literature search methodology, please refer to \tableautorefname~\ref{tab:keywords_table}.

\begin{figure}[t]
    \centering
    \includegraphics[width=\textwidth]{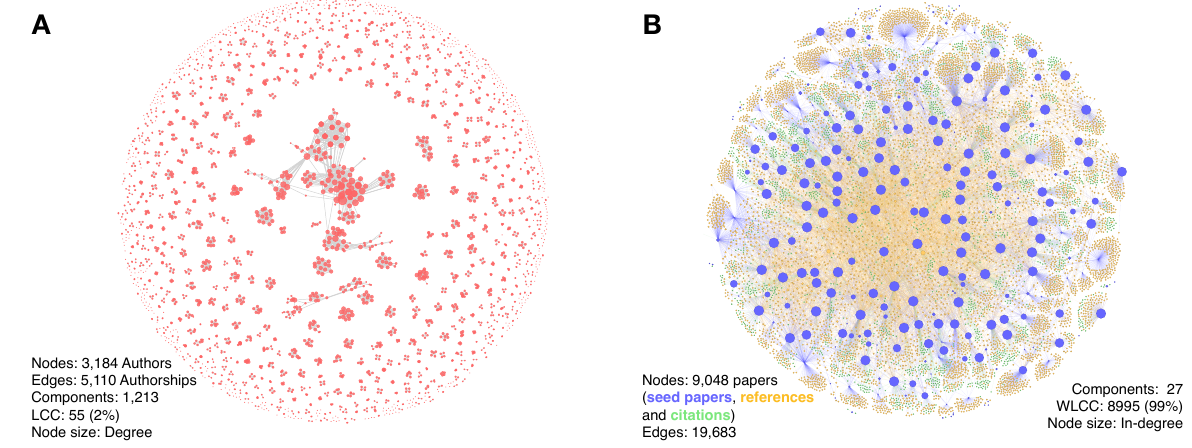}
    \caption{
        (\textbf{A}) Co-authorship network of the papers with keywords related to ``Complex Systems'' and ``Ethics''. This undirected network is made of 3,184 authors and 5,110 edges between them. The size of the nodes represents the co-authorships of the corresponding author. Notice how disconnected the network is, counting the 1,213 connected components, with the largest connected component (LCC) having just 55 nodes.
        (\textbf{B}) Citation network of the papers mentioning ``Complex Systems'' and ``Ethics'', including their references and citations. This directed network comprises 9,048 papers and 19,683 edges between them. The size of a node is proportional to the in-degree of the related paper, i.e., the number of papers citing the paper. Notice how the selected papers share similar citations, resulting in only 27 weakly connected components. The color of the nodes refers to the type of paper as written in the legend, and there are no overlapping categories. Then, if a paper that is seed paper is the reference of another paper, this stays as seed paper.
    }
    \label{fig:figure2}
\end{figure}

Additionally, we analyzed the co-authorship and citation networks of the retrieved literature to assess the cohesion among authors in terms of their collaborative work and common references. While the discussion of ethical challenges in complexity science has been limited, understanding the structure of the research community is crucial for addressing these gaps effectively. The co-authorship network is notably sparse, featuring many disconnected components, with its largest connected component (LCC) comprising only 2\% of the entire network (\figureautorefname~\ref{fig:figure2}\textbf{A}). In contrast, the citation network---which includes the retrieved papers, their references, their citations, and the connections among them---displays a more cohesive structure, with just 27 connected components and its LCC encompassing 99\% of the network (\figureautorefname~\ref{fig:figure2}\textbf{B}). These findings reflect the broader issue identified in the previous sections: the ethical discourse in complexity science is not only limited but also fragmented. This underscores the need for creating shared spaces, both online and in-person, to foster greater connections and collaborative research around these topics. By addressing these networking challenges, we can better facilitate discussions and development of ethical guidelines within the complexity science community.

\section*{Practical considerations for ethical actions in complexity science}

Given the interdisciplinary nature of the community of complexity scientists, which includes researchers from diverse fields, the ethical concerns we face are complex and multifaceted~\cite{balsamo2010interdisciplinarity}. To address these challenges, we propose to start the discussion by examining ethical frameworks and practices already established in other adjacent fields, identifying relevant aspects, and tailoring them to fit our specific needs. 
Ethical aspects raised by such adjacent fields include, but are not limited to, the following points:
\begin{itemize}
    \item \textit{Transparency}: ensuring openness in all stages of the research process, from data collection to the deployment of the results.
    \item \textit{Integrity}: maintaining honesty and responsibility in the research process, ensuring reliable contributions, and avoiding shortcuts or intellectual laziness. 
    \item \textit{Accountability}: establishing responsibility for the future use and potential consequences of the research findings.
    \item \textit{Fairness}: advocating for equity in how models affect different groups, avoiding biases.
    \item \textit{Minimal risk}: protecting the well-being of individuals and animals involved in research studies.
    \item \textit{Data privacy}: safeguarding sensitive information and maintaining confidentiality when using individual-level data.
    \item \textit{Working conditions}: ensuring ethical treatment of crowd workers involved in research.
\end{itemize}

These ethical concerns have prompted a few solutions implemented in other fields. Innovative examples of solutions in data privacy include personal online data repositories (PODs) and decentralized autonomous organizations (DAOs), which are emerging as secure, user-controlled, and democratic solutions for providing the industry with data while safeguarding users' privacy and including them in the value chain. These solutions are emerging in sectors such as healthcare~\cite{kpmg2023health,faruk2021towards}, research and education~\cite{filipvcic2022web3}, supply chain~\cite{ducree2020open}, and finance~\cite{dai2021flexible}, among others. 

Some traditional, institutionalized initiatives are also significant, including the Institutional Review Boards (IRBs), which are common in health, biological, and social sciences, and the Data Ethics Framework, which is implemented by the UK public sector. Non-institutional and voluntary efforts, such as the Fairness, Accountability, and Transparency in Machine Learning (FAT/ML) initiative and Human Rights Impact Assessments (HRIAs), are also noteworthy in addressing ethical concerns.

Nevertheless, grasping the full scope of ethical considerations in complexity science is inherently challenging, due to the field’s association with strong uncertainty and emergent phenomena. 
This unpredictability means that not all ethical dilemmas can be anticipated in advance, necessitating an ethical framework that is both flexible and adaptive. While governance, regulation, and fairness must often be evaluated on a case-by-case basis~\cite{weissinger2021ai}, these considerations must be integrated into complexity research.

Based on these considerations, we propose the following call for actions for the complexity science community on how to make an inclusive and pivotal change in the community's consideration and response to ethics and ethical concerns in complexity science:

\begin{itemize} 
\item To support and enhance the ongoing ethical discussions, facilitating seminars and workshops on ethics within the community, extending beyond traditional conference settings. 
\item To commission an open-access journal series and an immediate call for papers focused on ethics in complexity science to drive field-shifting discussions and research. 
\item To finance the growth of research on this underrepresented yet fundamental topic and to develop co-produced ethical guidelines, in consultation with the complexity science community.
\end{itemize}

\section*{Conclusion}

In conclusion, we acknowledge that scientific knowledge is never entirely objective and value-free. Ethical considerations can challenge existing epistemological and ontological assumptions. Therefore, here we aim to highlight the importance of ``Ethics in Complexity Science'' as a crucial area to explore and integrate. 
We do not suggest that everything should be controlled or forbidden. Rather, the current lack of consideration for ethical issues demonstrates the inability to future-proof complexity as a field and the need to maximize its potential to address significant social challenges. 

Finally, we want to reaffirm that, more than proposing a new framework, we seek to stimulate dialogue among researchers, practitioners, and policymakers from diverse disciplines, encouraging a deeper understanding of the intricate ethical concerns inherent in complex systems. We hope to pave the way for innovative approaches to address such challenges and opportunities presented by the ever-evolving dynamics of our field.

\section*{Author contributions}
All authors contributed equally to the development and writing of this work. The order of authors listed is purely alphabetical and does not reflect the extent of individual contributions.

\section*{Acknowledgements}
The authors wish to acknowledge Fabrizio Li Vigni and Markus Christen for the useful comments on an earlier version of this work.

\section*{Funding statement}
O.A received salary support from the UK Prevention Research Partnership (Violence, Health, and Society; MR-VO49879/1), which is funded by the British Heart Foundation, Chief Scientist Office of the Scottish Government Health and Social Care Directorates, Engineering and Physical Sciences Research Council, Economic and Social Research Council, Health and Social Care Research and Development Division (Welsh Government), Medical Research Council, National Institute for Health and Care Research, Natural Environment Research Council, Public Health Agency (Northern Ireland), The Health Foundation, and Wellcome. Y.A thanks the University of Zurich and the Digital Society Initiative for (partially) financing this project. The views expressed in this article are those of the authors and not necessarily those of the UK Prevention Research Partnership or any other funder.

% \bibliography{references} 
\printbibliography

\clearpage
\section*{Annexus}

\begin{table}[!h]
    \centering
    \resizebox{0.8\columnwidth}{!}{
    \begin{tabular}{c|l}
        
         \textbf{Search label} & \textbf{Keywords} \\
         \hline
         &\\
         \multirow{18}{*}{\textbf{Ethics}} &  is any of (clinical-ethics, animal-ethics, marketing-ethics,  military-medical-ethics,\\
         & criminal-justice-ethics, computer-ethics, ethics-of-technology, applied-ethics, ethical-dilemmas,\\
         & christian-ethics, nuclear-ethics, ethics, selection-bias, algorithmic-bias, transparency,\\
         & accountability, surveillance, ethical-challenges, ethical-issues, ethical-dilemma, ethical-egoism,\\
         & discrimination, neuroethics, responsibility-in-ai, moral-judgment, fairness, ethical-considerations,\\
         & medical-ethics, healthcare-ethics, ethics-committee, ethics-consultation, ethics-and-leadership,\\
         & ethics-education, ethics-in-law, ethics-in-science, ethics-in-business, ethics-in-accounting,\\
         & ethics-of-courtroom-imagery, ethics-of-teaching, ethics-of-care, ethics-of-representation,\\
         & ethical-standards, ethical-code, ethical-theories, ethical-decision, ethical-identity,\\
         & ethical-trading-initiatives, ethical-data-collection, ethical-toolbox, ethical-and-legal-issues,\\
         & ethical-behavior, ethical-codes, ethical-consequences, ethical-concepts, ethical-climate,\\
         & ethical-consumption, ethical-competence, ethical-decision-making, ethical-dimensions,\\
         & ethical-education, ethical-guidelines, ethical-governance, ethical-implications, ethical-leadership,\\
         & ethical-lessons, ethical-principles, ethical-practices, ethical-responsibility, ethical-subjectivity,\\
         & ethical-sustainability, ethical-theory, ethical-values, bioethics, bioethical-issues, research-ethics,\\
         & virtue-ethics, environmental-ethics, normative-ethics, business-ethics, empirical-ethics, \\
         &corporate-ethics, legal-ethics, interpretability,  trustworthiness)\\
         & \\
         \hline
         & \\
         \multirow{7}{*}{\textbf{Ethics and AI}} &  Ethics* \textbf{AND} is any of (artificial-intelligences, machine-learning, deep-learning,\\
         & support-vector-machines, data-mining, fault-detection, big-data-analytics, big-data,\\
         & temporal-data-mining, temporal-data-analysis, image-recognition, adversarial-machine-learning, \\
         & machine-learning-algorithms, margin, online-machine-learning, machine-learning-attacks, \\
         & quantum-machine-learning, machine-learning-interpretability, automated-machine-learning,\\
         & data-analytics, data-privacy, interpretable-models, privacy-concerns, anticipation,\\
         & predictability, statistical-learning, model-interpretability)\\
         & \\
         \hline
         & \\
         \multirow{4}{*}{\textbf{Ethics and CompSys}} &   Ethics* \textbf{AND} is any of (complex-system, complexity-management, complexity-science, \\
         & network-analysis, social-networks, online-social-networks, agent-based-model,\\
         & agent-based-modeling, agent-based-simulation, multi-agent-systems, epidemic-spreading,\\
         & complexity-economics, complex-systems, complex-system-modeling, community-detection,\\
         &social-network-analysis)\\
         & \\
    \end{tabular}}
    \caption{Keywords used to retrieve literature from the Open Alex database \cite{openalex} until 2023. The keywords are divided into three categories: the row ``Ethics'' encompasses a broader amount of literature, and the other two rows consider the intersection between the ``Ethics'' literature and the literature in ``AI'' and ``Complex systems (CompSys)''. No search was done in the full text, abstract, or title.}
    \label{tab:keywords_table}
\end{table}

\end{document}